\documentstyle[proceedings,numreferences]{maxent95}
\makeindex                  

\begin{opening}
\title{MaxEnt Queries and Sequential Sampling}
\author{Peter Riegler}
\institute{ABB Corporate Research, Speyerer Stra{\ss}e 4, D-69115 Heidelberg,
Germany\footnote{Email: peter.riegler@de.abb.com}}
\author{Nestor Caticha}
\institute{Instituto de F\'{\i}sica, Universidade de S\~{a}o Paulo, CP
66318 S\~{a}o Paulo, SP, CEP05389-970 Brazil\footnote{Email: nestor@if.usp.br}}
\end{opening}
\begin{document}
\begin{abstract}
In this paper we pose the question: After gathering $N$ data points, at
what value of the control parameter should the next measurement be done? We
propose an on-line algorithm which samples optimally by maximizing the gain
in information on the parameters to be measured. We show analytically that
the information gain is maximum for those potential measurements whose
outcome is most unpredictable, i.e.\ for which the predictive distribution
has maximum entropy. The resulting algorithm is applied to exponential
analysis.
\end{abstract}

\section{Introduction}
Experiments usually measure the response of a physical system to an input.
The measured set of outputs together with the 
corresponding inputs is then used to determine the 
parameters of some assumed model. For instance, for determining the
electric resistance one would measure the resulting current for several
values of the applied voltage. The voltage acts as a sort of control
parameter. For time series the corresponding control parameter would be the
time instances at which the signal is sampled. 

The most common procedure for measuring physical parameters is to sample the
underlying physical signal equidistantly with respect to the control
parameter. However, it is often not clear how the accuracy of the parameters
to be measured depends on the range over which the control parameter is
varied. This is typical for signals of the form $\exp(-t/\tau)$, which decay
exponentially in the control parameter $t$. For instance, if one happens to
choose the maximum value of the control parameter to be much larger than the
unknown decay constant $\tau$ many samples would obviously be ``pure
noise'', thus reducing the accuracy of the estimate of $\tau$.

The situation is particularly severe for sums of exponentials. There, the
sampling rate clearly determines the smallest resolvable decay constant.
Since for \index{equidistant sampling} equidistant sampling  
those parts of the signal 
with larger decay
constants are sampled more often, the estimates become the more accurate the
larger the decay constant. Hence, a lot of 
sampling is wasted in order to 
estimate the easily to determine larger decay constants.

An experimentalist measuring a spectrum of decay constants consequently
faces the question how to choose the values of the control parameters in
order to obtain accurate measurements while restricting the
maximum number of experiments to $N$. This task is basically one of 
\index{design of experiment} design of experiment, 
a discipline well-known in statistics \cite{Tagouchi}.
Due to the nonlinear dependence on the decay constants this is a nontrivial
task, however. It would involve an optimization with respect to the control
parameters of the $N$ measurements to be done.

In this paper, we take a simplified approach by posing
the question: After gathering $N$ data points, at
what value of the control parameter should the next measurement be done? We
propose an on-line algorithm which samples optimally by maximizing the gain
in information on the parameters to be measured.  By this, the
experiment is designed ``sequentially'' or ``on-line''
rather than completely from the onset.

Due to its design, the algorithm will choose at each step of the experiment
the value of the control parameter at which the next experiment should
be performed. In Section \ref{secMaxEnt} we choose 
the entropy of the posterior as an
information measure and introduce the algorithm for sequential
\index{sequential MaxEnt sampling} MaxEnt sampling. 
We show analytically that the algorithm reduces to
finding that value of the control parameter for which the width of the
predictive distribution is maximum.

Section \ref{secMonoexp}
gives an application of the algorithm to exponentially decaying signals
and Section \ref{secEquidistant} compares the performance of the algorithm with
equidistant sampling. In Section \ref{secMultiexp} we apply the algorithm
to exponentially decaying signals having more than one decay constant.
Finally, Section \ref{secSummary} summarizes the results and provides a
conclusion.

\section{MaxEnt formulation of sequential sampling\label{secMaxEnt}}

Assume we have determined a discrete data set 
$D_N = \{(t_1,s_1),...,(t_N,s_N)\}$,
sampled from $s(t)$ at discrete instances of the control parameter
$t_1,...,t_N$, with a model equation
\begin{equation}
\label{signal}
s_i = s(t_i) = f(t_i) + \xi_i,
\end{equation}
where $f(t)$ is the signal and $\xi$ represents noise in the problem.
Usually one has some parameterized form of $f(t)$ and the aim of the
experiment is to determine (at least some of) these parameters. Having
obtained $D_N$, the knowledge of these parameters is contained in the
posterior distribution
\begin{equation}
P(\mbox{\boldmath $\beta$\unboldmath}|D_N)= 
\frac{P(D_N|\mbox{\boldmath $\beta$\unboldmath}) 
P(\mbox{\boldmath $\beta$\unboldmath})}{P(D_N)}
,
\end{equation}
where $\mbox{\boldmath $\beta$\unboldmath}$ denotes the whole set of
parameters to be determined.  For the prior probability of the noise
$P(\xi)=P(D_N|\mbox{\boldmath $\beta$\unboldmath})$ we take the
least informative prior \cite{Bretthorst88}, 
i.e. a Gaussian of zero mean and variance
$\sigma^2$.

The aim of the experiment is to maximize
the information about $\mbox{\boldmath $\beta$\unboldmath}$ during
the course of the experiment. Using the negative entropy of the posterior as a
measure of information this amounts to minimizing the entropy
\begin{equation}
\label{SDN}
S(D_N) = - \int d\mbox{\boldmath $\beta$\unboldmath}
P(\mbox{\boldmath $\beta$\unboldmath}|D_N)
\log P(\mbox{\boldmath $\beta$\unboldmath}|D_N)
.
\end{equation}
The most informative experiment would then be defined by those
values of the control parameter $t$ which satisfy $S(D_n)= min$, implying
\begin{equation}
\label{fullyoptimized}
\frac{\partial S(D_N)}{\partial{t_k}}=0
\mbox{~~~for all~~~} k=1,...,N
.
\end{equation}
In general this equation will be too difficult to solve.
Moreover, the optimization (\ref{fullyoptimized}) is not 
advisable, since it relies on a fixed model for $f(t)$.
In general, there might be different optimal experiments for different
models.

These difficulties can be avoided if one chooses to design the 
experiment in a sequential way.
Instead of determining the values of all $t_i$ one could  
ask the question: After gathering $N$ data points, at
what value of the control parameter $t_{N+1}$ should the next measurement 
be done in order to gain as much information as possible about the parameters
$\mbox{\boldmath $\beta$\unboldmath}$ to be estimated? 

The answer to this
is obtained by varying $S(D_{N+1})=S\left(D_N \cup (t_{N+1},s_{N+1})\right)$ 
with respect to the value of the control parameter $t_{N+1}$ 
at the next experiment.
That is, one has to determine the minimum of 
\begin{equation}
\label{Stq}
\overline{S}(t_{N+1}) = \int ds_{N+1} S(D_{N+1}) P(s_{N+1}|D_N,t_{N+1})
\end{equation}
with respect to $t_{N+1}$. Note that one has to average out the 
unknown outcome $s_{N+1}$ of the next experiment using 
the predictive distribution \index{predictive distribution}
$P(s_{N+1}|D_N,t_{N+1})$.
The predictive reflects the 
knowledge about $s_{N+1}$ given the previously obtained data $D_N$
and the control parameter of the next experiment.

Our sequential approach to design optimal experiments is somewhat
similar to so-called query \index{query} learning in neural networks.
There one searches for such training examples that speed up the
learning process considerably 
\cite{KinzelRujan1990,SollichThesis,Fukumizu2000}.
Based on this similarity 
we name the value of the control parameter $t_{N+1}$ which minimizes 
(\ref{Stq}) the {\em query} $t_q$.

Applying Bayes rule one can rewrite the r.h.s.\ of (\ref{Stq}) in the form
\begin{eqnarray}
\label{fullentropy}
\overline{S}(t_{N+1}) &=&
\int ds_{N+1} P(s_{N+1}|D_N,t_{N+1}) \log P(s_{N+1}|D_N,t_{N+1})
\nonumber \\
&&-\int d\mbox{\boldmath $\beta$\unboldmath}
P(\mbox{\boldmath $\beta$\unboldmath}|D_N)
\log P(\mbox{\boldmath $\beta$\unboldmath}|D_N)
+\frac{1}{2}\left(1+\log2\pi\sigma^2\right)
.
\end{eqnarray}
Note that the only dependence on $t_{N+1}$ is via
the entropy of the predictive distribution
\begin{equation}
S_{pred}(\tau)= -\int ds_{N+1} P(s_{N+1}|D_N,\tau) \log P(s_{N+1}|D_N,\tau)
.
\end{equation}
Hence, in order to minimize the entropy of the posterior with a new
measurement $(t_q,s(t_q))$ and $N$ old measurements $D_N$, one has to
{\em maximize} the entropy of the predictive distribution.
The reason for this is that $t_q$ is that value of the control parameter
for which the prediction is least secure. Since the derived sampling 
procedure is based on a maximum entropy criterion, we call it
sequential MaxEnt sampling.

Two notes are in place here, concerning the implementation of the
above algorithm: According to (\ref{fullentropy}), only 
$P(s_{N+1}|D_N,t_{N+1})$ is needed in order to determine $t_q$.
By definition
\begin{equation}
\label{defpred}
P(s_{N+1}|D_N,t_{N+1})=\frac{P(D_{N+1})}{P(D_{N})}
=\frac{\int d\mbox{\boldmath $\beta$\unboldmath}
P(\mbox{\boldmath $\beta$\unboldmath}|D_{N+1})}
{\int d\mbox{\boldmath $\beta$\unboldmath}
P(\mbox{\boldmath $\beta$\unboldmath}|D_N)}
.
\end{equation}
For  model functions of the form
\begin{equation}
\label{genmodel}
f(t) = \sum_{j=1}^{m} A_j G_j(t,\{\lambda\})
\end{equation}
the integration with respect to the model parameters 
$\mbox{\boldmath$\beta$\unboldmath}=A_j,\lambda,\sigma$
can be simplified if one treats 
the parameters $A_j$ as nuisance parameters, see \cite{Bretthorst88}
for details. In the same manner, the variance of the noise in 
(\ref{signal}) can be integrated out analytically using Jeffreys prior. 
Hence, the only parameters that actually need to be integrated over
in (\ref{defpred}) are the $\lambda$, which enter $f(t)$ in a nonlinear
manner, in general. 

\begin{figure}
\begin{center}
\setlength{\unitlength}{1pt}
\begin{picture}(325,115)(0,0)
 \put(0,0){\makebox(325,115)
          {\includegraphics{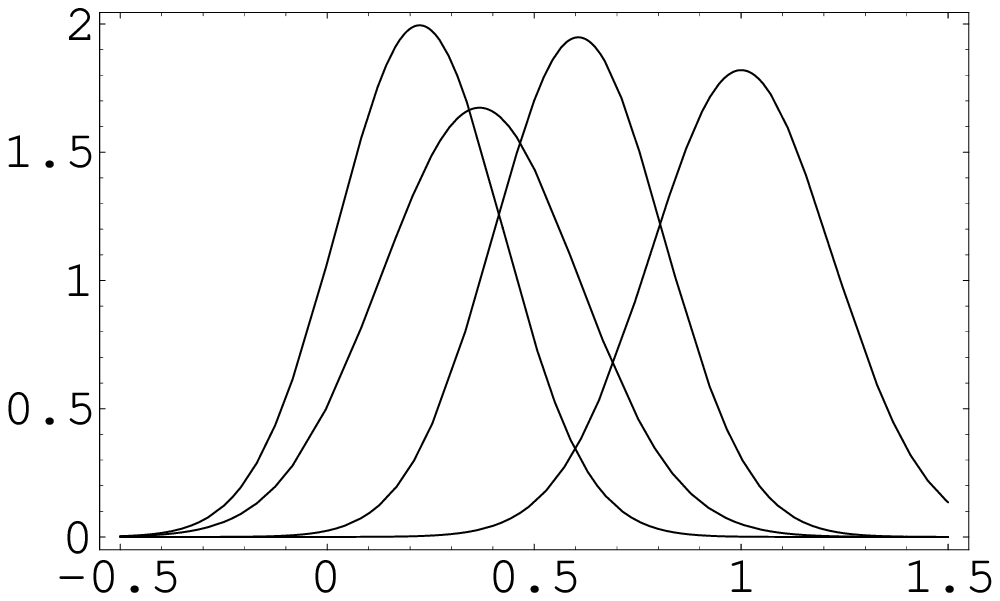}}}
 \put(0,0){\makebox(325,115)
          {\includegraphics{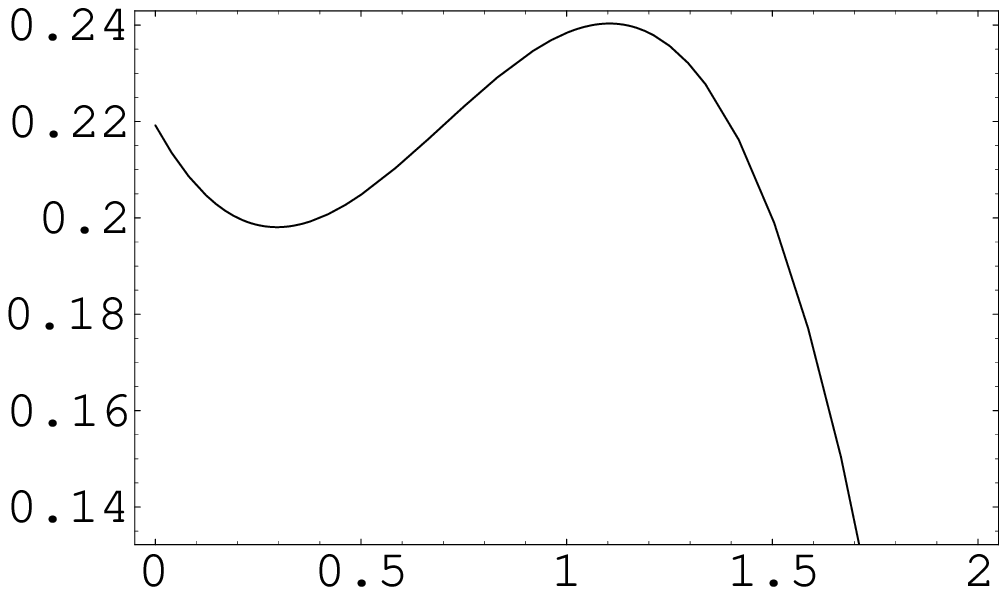}}}
\put(165.00,52.50){\makebox(0,0)[lc]{$S_{pred}$}}
\put(80.00,-10.00){\makebox(0,0)[lc]{$s_{N+1}$}}
\put(255.00,-10.00){\makebox(0,0)[lc]{$t_{N+1}$}}
\end{picture}
\end{center}
\caption{Left:
Predictive distribution for various values of $t_{N+1}$ (left to right:
$t_{N+1}=$ 1.5, 1.0,0.5,0). In the case shown the
next measurement would be performed at $t_q=1.1$. For this value
the predictive distribution has the largest width and the corresponding
entropy is maximum (right). The results have been obtained for a
monoexponentially decaying signal as discussed in Section \ref{secMonoexp}. 
}
\label{fig_predictive}
\end{figure}

For sufficiently large values of $N$ the predictive distribution can
be well approximated by a Gaussian of width $\sigma_{pred}$. The most
probable value of $s(t_{N+1})$ will then be 
$(2 \pi \sigma_{pred}^2)^{-1/2}$,
which depends on $t_{N+1}$ via  $\sigma_{pred}$. Maximization of $S_{pred}$
then amounts to searching for that value of  $t_{N+1}$ for which the
most probable value of $s(t_{N+1})$ is minimum, 
see Figure \ref{fig_predictive}.

\section{Application to a monoexponentially decaying signal\label{secMonoexp}}

As a first example we apply the proposed sequential MaxEnt sampling
procedure to an exponentially decaying signal with only one decay
constant:
\begin{equation}
\label{monoexponential}
s(t) = A \exp(-\lambda t) + \xi
.
\end{equation}
As before, $\xi$ denotes additive noise of variance $\sigma$.

It is obvious that the sampling technique proposed in Section \ref{secMaxEnt}
cannot be applied from scratch, i.e. without any experimental data at hand.
For estimating $\lambda$, for instance, one needs to have at least two
measurements at different time instances since one needs at least two 
points to fit a straight line to $\log s$. 
Here we use three initial measurements, because 
two measurements lead to a posterior of width 0 which is difficult to 
handle numerically.

\begin{figure}
\begin{center}
\setlength{\unitlength}{1pt}
\begin{picture}(325,115)(0,0)
 \put(0,0){\makebox(325,115)
          {\includegraphics{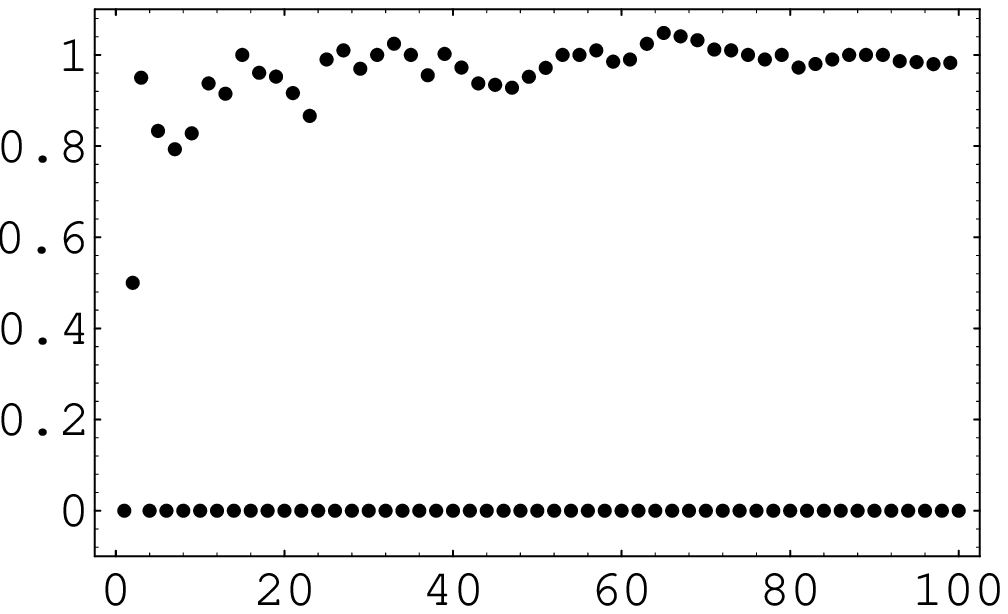}}}
 \put(0,0){\makebox(325,115)
          {\includegraphics{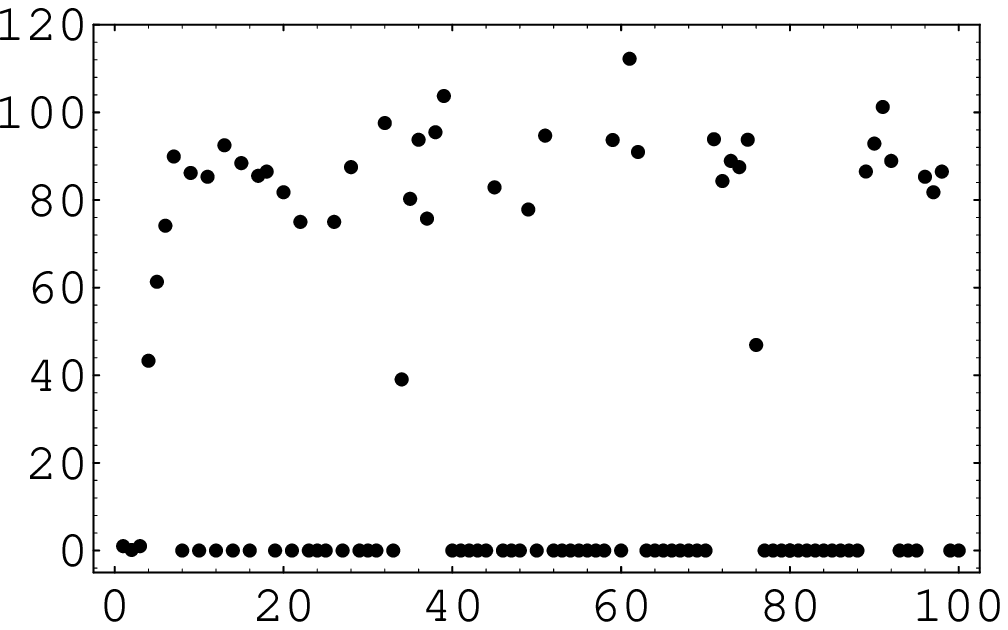}}}
\put(165.00,50.00){\makebox(0,0)[lc]{$t_q$}}
\put(-10.00,50.00){\makebox(0,0)[lc]{$t_q$}}
\put(80.00,-10.00){\makebox(0,0)[lc]{$N$}}
\put(255.00,-10.00){\makebox(0,0)[lc]{$N$}}
\end{picture}
\end{center}
\caption{Simulation results for the sampling times $t_q$ for $A=1$,
$\sigma=0.1$. The decay constant has been chosen to be $\lambda=1$ (left)
and $\lambda=0.01$ (right)
}
\label{fig_tq}
\end{figure}

For the simulation results shown in the following, the required
three initial measurements have been generated by drawing $t_i$, $i=1..3$ 
from a uniform distribution in $[0,2]$. 
In general, one would perform the initial measurements
within an interval which is of the order of the decay
time $\tau=1/\lambda$. Note that this choice should be possible in
principle, since an experimentalist will have some prior knowledge about
the decay constant. However, such a choice is not
mandatory, as is exemplified in Figure \ref{fig_tq}.
There we show simulation
results for  $\lambda=1$ and $\lambda=0.01$. 
In the latter case
the initial measurements are at times 
much smaller than the decay constant $\tau=100$, yet the sequential
MaxEnt sampling algorithm is capable of tracking the signal.

Figure \ref{fig_tq} shows the evolution of sample times $t_q$ for
increasing length $N$ of the experiment. We find the remarking behavior
that the sampling times ``oscillate'' between $t_q=0$ and $t_q \simeq
1/\lambda$. Note that the period is not strictly equal to one, since
sometimes two or more consecutive measurements are at either $t_q=0$ or $t_q
\simeq 1/\lambda$. The oscillating behavior can be traced back to the
functional form of $S_{pred}(t_{N+1})$, cf.\ Figure \ref{fig_predictive}. 
The entropy $S_{pred}(t_{N+1})$ has two
maxima for $t \geq 0$, the absolute maximum determining $t_q$.
By construction of MaxEnt sampling the next measurement will be placed at
$t_{global}$ and $S(t_{global})$ will decrease relative to $S(t_{local})$. 
After one or more measurements the predictive entropy 
\index{predictive entropy} at the hitherto global 
maximum will be lower than at the hitherto local maximum and the
two maxima will change their roles.

\begin{figure}
\begin{center}
\setlength{\unitlength}{1pt}
\begin{picture}(325,115)(0,0)
 \put(0,0){\makebox(325,115)
          {\includegraphics{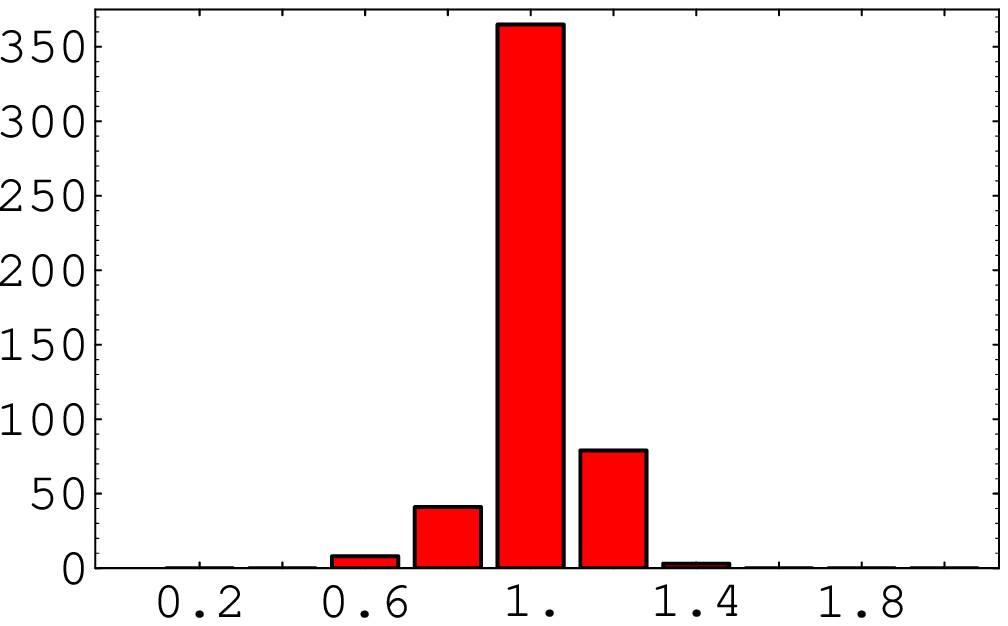}}}
 \put(0,0){\makebox(325,115)
          {\includegraphics{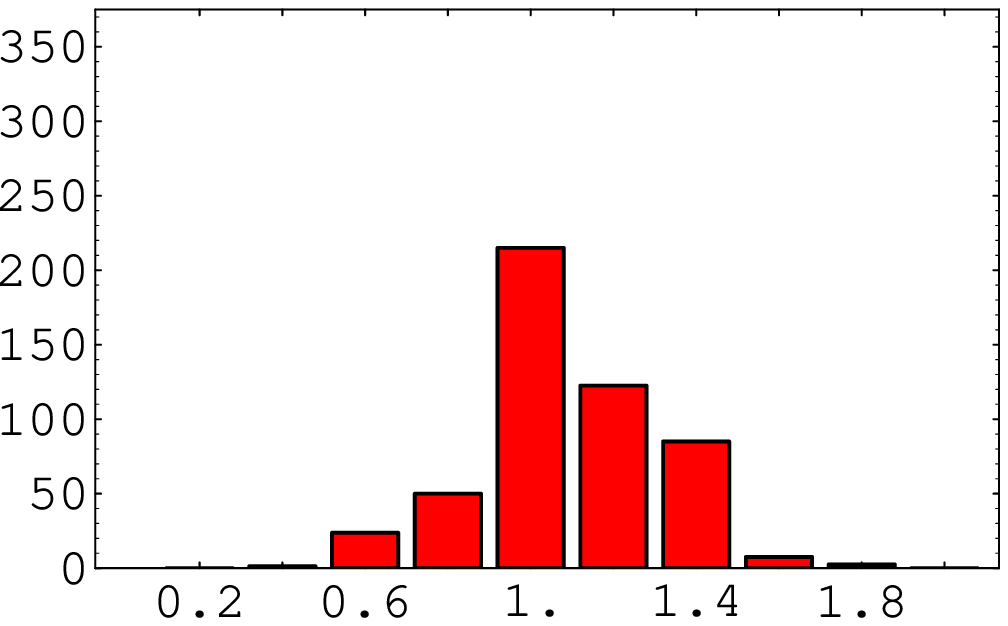}}}
\put(80.00,-10.00){\makebox(0,0)[lc]{$t_q$}}
\put(255.00,-10.00){\makebox(0,0)[lc]{$t_q$}}
\end{picture}
\end{center}
\caption{Distribution of sampling times $t_q \neq 0$ for an exponentially
decaying signal of form (\ref{monoexponential}) with $A=1$ and
$\lambda=1$.  The histograms have been obtained for 10 experiments of
length $N=100$. Left: $\sigma=0.1$, right: $\sigma=0.5$.
}
\label{fig_histo}
\end{figure}

From Figure \ref{fig_tq} it is evident that the larger sampling times are
not exactly at $t_q=1/\lambda$. Instead, the values of $t_q$ have some
spread which increases with the variance $\sigma^2$ of the additive noise, 
see Figure \ref{fig_histo}.

\begin{figure}[b]
\begin{center}
\setlength{\unitlength}{1pt}
\begin{picture}(325,115)(0,0)
 \put(0,0){\makebox(325,115)
          {\includegraphics{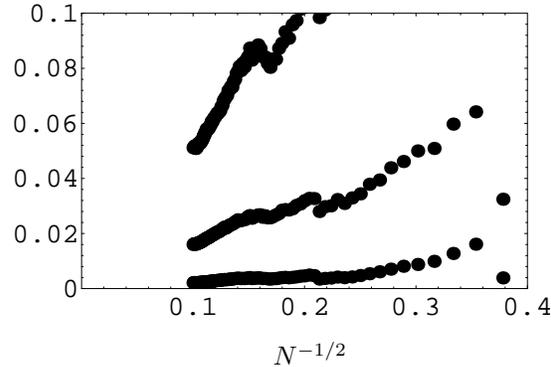}}}
\put(170.00,-5.00){\makebox(0,0)[lc]{$N^{-1/2}$}}
\end{picture}
\end{center}
\caption{Standard errors of the estimates of 
$\lambda$, $A$, and $\sigma$ (top to bottom)
obtained via sequential MaxEnt sampling. Shown are the results for the same
simulation as in Figure \ref{fig_tq}, left.
}
\label{fig_asymptotics}
\end{figure}

In the course of the experiment
the standard errors of the parameters to be estimated decrease with
$1/\sqrt{N}$, asymptotically, as expected, cf.\ \cite{Bretthorst88}. 
See Figure \ref{fig_asymptotics} for an example.

A note is in place here concerning the oscillating behavior of the
control parameter. For time series, the control parameter $t$ would be
time. At first glance, it might seem impossible to perform sequential
MaxEnt sampling since time can only be increased but not decreased. For 
many types of experiments, however, one can actually decrease the time
of the next measurement by performing a new experiment. For instance, in
measuring NMR relaxation times one would first magnetize the probe 
which defines the onset of the experiment, $t=0$, and then measure the
magnetization for several $t>0$. For MaxEnt sampling, one would have
to magnetize, measure at $t_q$, magnetize again, and so on. Hence,
$t_q$ can increase or decrease in consecutive experiments.

\section{Comparison with equidistant sampling\label{secEquidistant}}

It is instructive to compare the results obtained for sequential MaxEnt
sampling with other sampling methods. The most common alternative method
is probably equidistant sampling where the experiments are performed at $N$
equidistantly spaced values of the control parameter. For a signal of form
(\ref{monoexponential}) the lowest of these values is obviously $t=0$ while
the largest  $t_{max}$ is up to the choice of the experimentalist.

\begin{figure}
\begin{center}
\setlength{\unitlength}{1pt}
\begin{picture}(325,115)(0,0)
 \put(0,0){\makebox(325,115)
          {\includegraphics{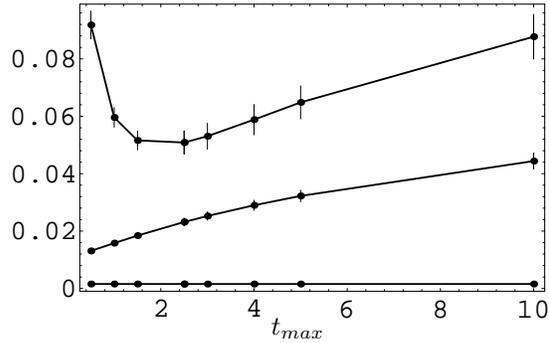}}}
\put(170.00,-5.00){\makebox(0,0)[lc]{$t_{max}$}}
\end{picture}
\end{center}
\caption{Standard errors of the  Bayesian estimates of 
$A$, $\lambda$, $\sigma$ 
of a signal of form (\ref{monoexponential}).
Results have been obtained for equidistant sampling of $N=100$ points between
$t=0$ and $t_{max}$, averaged over 20 independent
runs (same simulation settings as in Figure \ref{fig_tq}, left). 
Top to bottom: standard errors of $\lambda$, $A$, $\sigma$.
Note that depending on the parameters to be estimated 
there is a different optimal choice for $t_{max}$.
The standard error of $\sigma$ is basically independent of $t_{max}$.
}
\label{fig_equi}
\end{figure}

Clearly, there is an optimal choice of $t_{max}$. For, choosing 
$t_{max} \gg 1/\lambda$ would result in sampling of noise only, while
$t_{max} \ll 1/\lambda$ would give a poor estimate of $\lambda$ as the
noiseless signal $A \exp(-\lambda t)$ hardly varies in this range. This
behavior is confirmed by simulations. In Figure  \ref{fig_equi} we show
the standard errors of the Bayesian estimates of $A$, $\lambda$, $\sigma$
for various values of $t_{max}$. As can be seen, the standard error of $A$
increases with $t_{max}$ while the standard error of $\lambda$ has some
minimum around $t_{max}\simeq2/\lambda$; the standard error of the estimate
of $\sigma$ is basically independent of $t_{max}$. 

Consequently, in choosing $t_{max}$ there is a tradeoff between estimating
$A$ (leading to $t_{max}\to 0$) and $\lambda$ 
(leading to $t_{max}\simeq 2/\lambda$). Besides this tradeoff, not knowing
$\lambda$ one could determine $t_{max}$ only based on the prior knowledge on
$\lambda$. In contrast to that, sequential MaxEnt sampling, as described in
the previous section, does not require the setting of any maximum control
parameter $t_{max}$ and places its measurements automatically.

\begin{figure}
\begin{center}
\setlength{\unitlength}{1pt}
\begin{picture}(325,115)(0,0)
 \put(0,0){\makebox(325,115)
          {\includegraphics{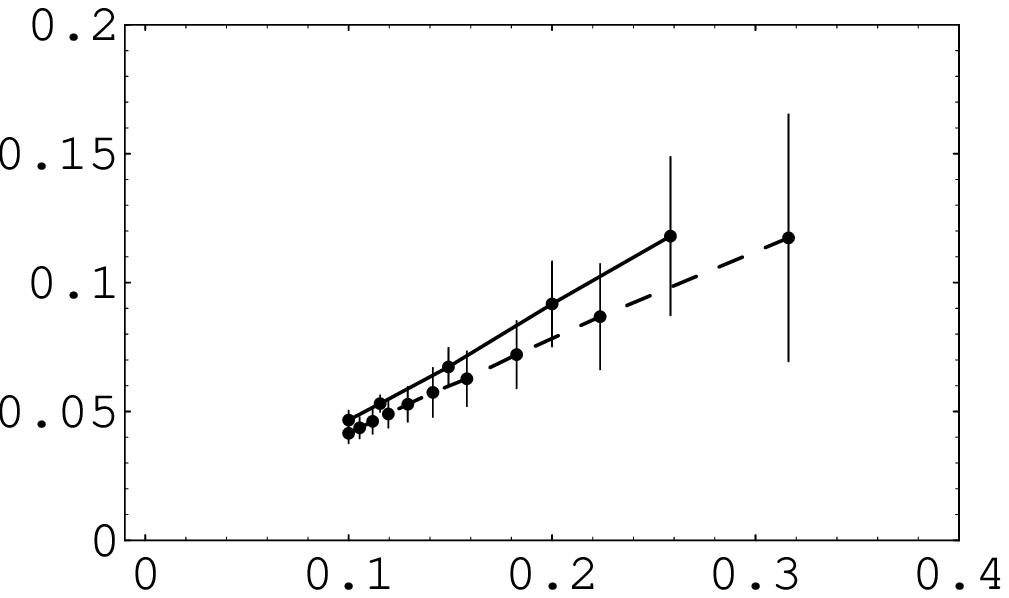}}}
 \put(0,0){\makebox(325,115)
          {\includegraphics{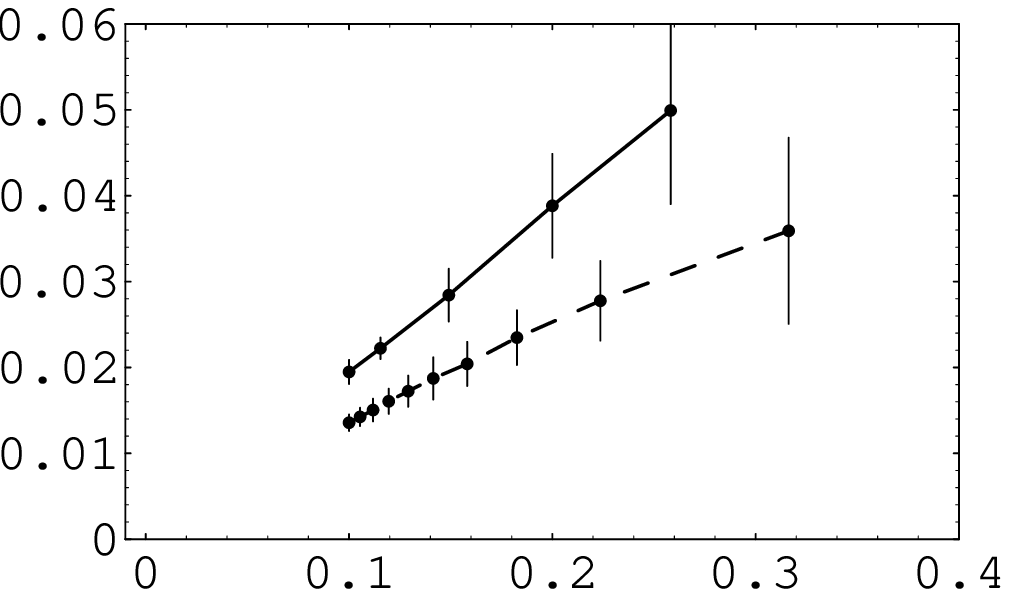}}}
\put(80.00,-10.00){\makebox(0,0)[lc]{$N^{-1/2}$}}
\put(255.00,-10.00){\makebox(0,0)[lc]{$N^{-1/2}$}}
\end{picture}
\end{center}
\caption{Comparison between sequential MaxEnt sampling (dashed line)
and equidistant sampling (solid line). Shown are the standard errors for
$\lambda$ (left) and $A$ (right). Simulation parameters are for 
the same settings as in Figure \ref{fig_equi}, averaged over 20
independent runs. Equidistant sampling has been performed with
$t_{max}=2.0$.
}
\label{fig_comparison}
\end{figure}

In addition, MaxEnt sampling leads to a much better resolution of the
parameters to be estimated. This is exemplified in 
Figure \ref{fig_comparison}, which compares the obtained standard errors
of $A$, $\lambda$ for MaxEnt and equidistant sampling.

\section{Application to sums of exponentials\label{secMultiexp}}
Exponentially decaying signals can be found in many physical phenomena as 
first order differential equations are among the most common in physics. 
However, the spectrum $\{\lambda_n,A_n\}$ of a multi-exponential
signal \index{multi-exponential signal}
\begin{equation}
\label{multiexponential}
f(t) = \sum_n A_n \exp(-\lambda_n t)
\end{equation}
is difficult to estimate. The reason is that the required inverse (discrete)
Laplace transform \index{Laplace transform} 
is known to be an ill-posed problem \cite{Huepper99,Vapnik}.
A small amount of noise added to $f(t)$ leads to a nonvanishing contribution
to the spectrum even in the limit of a very small perturbation.

Moreover, the quality of the spectrum to be estimated from measurements
depends strongly on the values of the control parameter $t$ at which the 
measurements are performed \cite{Istratov1999}. For instance, choosing
$t>1/\lambda^*$ certainly rules out the detection of any $\lambda_n>\lambda^*$ 
in (\ref{multiexponential}). Some good prior knowledge about the values of
$\lambda_n$ to be expected in the signal would be very helpful. But even
such knowledge would leave open the question how to choose the sampling
times $t_i$. For instance, in porous materials \cite{Rabbani1994}
the magnetization decays approximately like 
\begin{equation}
\label{porous}
f(t)=\sum_{n=1}^{\infty} \frac{A_0}{n^2} \exp\left(-\lambda_0 n^2 t\right)
,
\end{equation}
i.e. relaxation times and corresponding amplitudes rapidly decrease like
$1/n^2$. How does one have to choose the sampling times in order to
resolve the spectrum up to a desired $n_{max}$?

\begin{figure}
\begin{center}
\setlength{\unitlength}{1pt}
\begin{picture}(325,115)(0,0)
 \put(0,0){\makebox(325,115)
          {\includegraphics{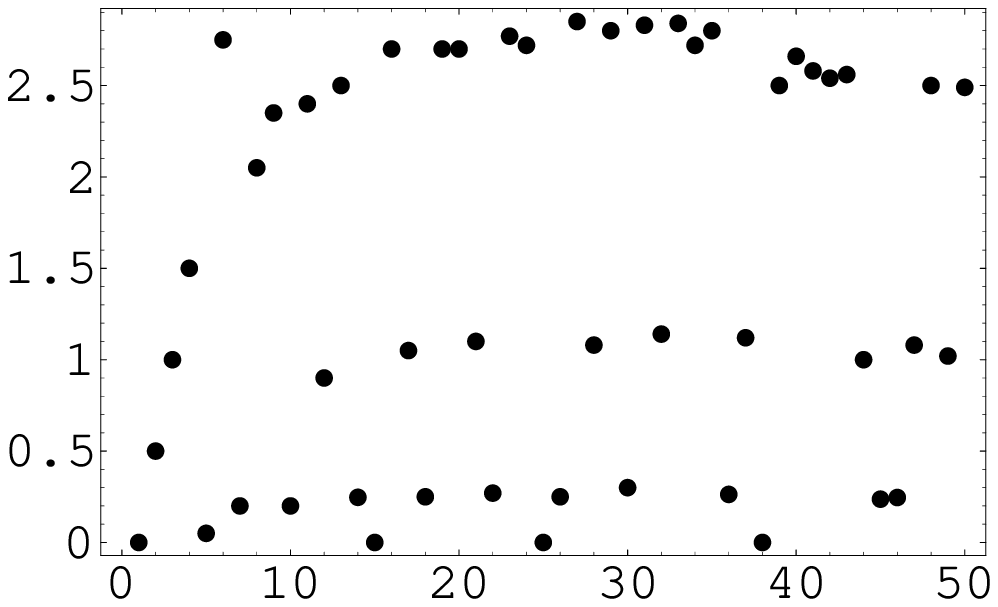}}}
 \put(0,0){\makebox(325,115)
          {\includegraphics{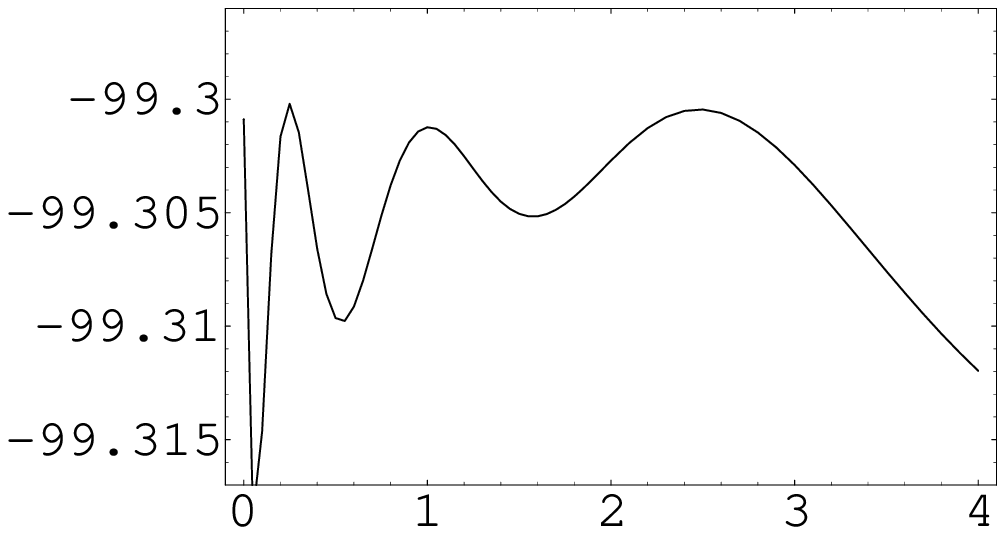}}}  
\put(165.00,68.00){\makebox(0,0)[lc]{$S_{pred}$}}
\put(-12.00,68.00){\makebox(0,0)[lc]{$t_q$}}
\put(70.00,-10.00){\makebox(0,0)[lc]{$s_{N+1}$}}
\put(255.00,-10.00){\makebox(0,0)[lc]{$t_{N+1}$}}
\end{picture}
\end{center}
\caption{
Left:
Simulation results for the sampling times $t_q$ for a signal of
type (\ref{multiexponential}) with two exponentials. 
Parameters are $A_1=1$, $A_2=1/4$,
$\lambda_1=1$, $\lambda_2=4$, $\sigma=0.01$.
Note that the first four measurements are chosen equidistantly in
order to start up the experiment.
Right: Entropy of the corresponding predictive distribution after $N=50$
measurements.
}
\label{fig_multitq}
\end{figure}

By construction of the algorithm, sequential MaxEnt 
sampling provides the answer to this question. The choice of $n_{max}$
fixes the model function (\ref{multiexponential}). Given some initial
data, the best query $t_q$ can be computed as described in Section 
\ref{secMaxEnt}. Figure \ref{fig_multitq} shows the  simulation result for the
values of $t_q$ for a signal of form (\ref{porous}) with $n_{max}=2$.

As in the monoexponential case, the values of the sampled control
parameters oscillate between $t=0$ and the characteristic time constants
in the signal, i.e. $1/\lambda_1$ and $1/\lambda_2$ in our case. Note
that in Figure \ref{fig_multitq} the  signal is also sampled at times 
$t > 1/\lambda_1$ larger
than the largest characteristic time constant 
The reason for this is probably due to the fact
that we have not imposed any priors
on the parameters $\lambda$. In order to rule out the presence
of any $\lambda<0$ one would clearly have to sample at large
values of $t$.

Figure \ref{fig_multitq} also depicts the characteristic
dependence of the predictive entropy on $t_{N+1}$. As in Section
\ref{secMonoexp} the location of its maxima characterizes the
sampled values of the control parameter $t$.

\begin{figure}
\begin{center}
\setlength{\unitlength}{1pt}
\begin{picture}(325,180)(0,0)
 \put(0,0){\makebox(325,180)
          {\includegraphics{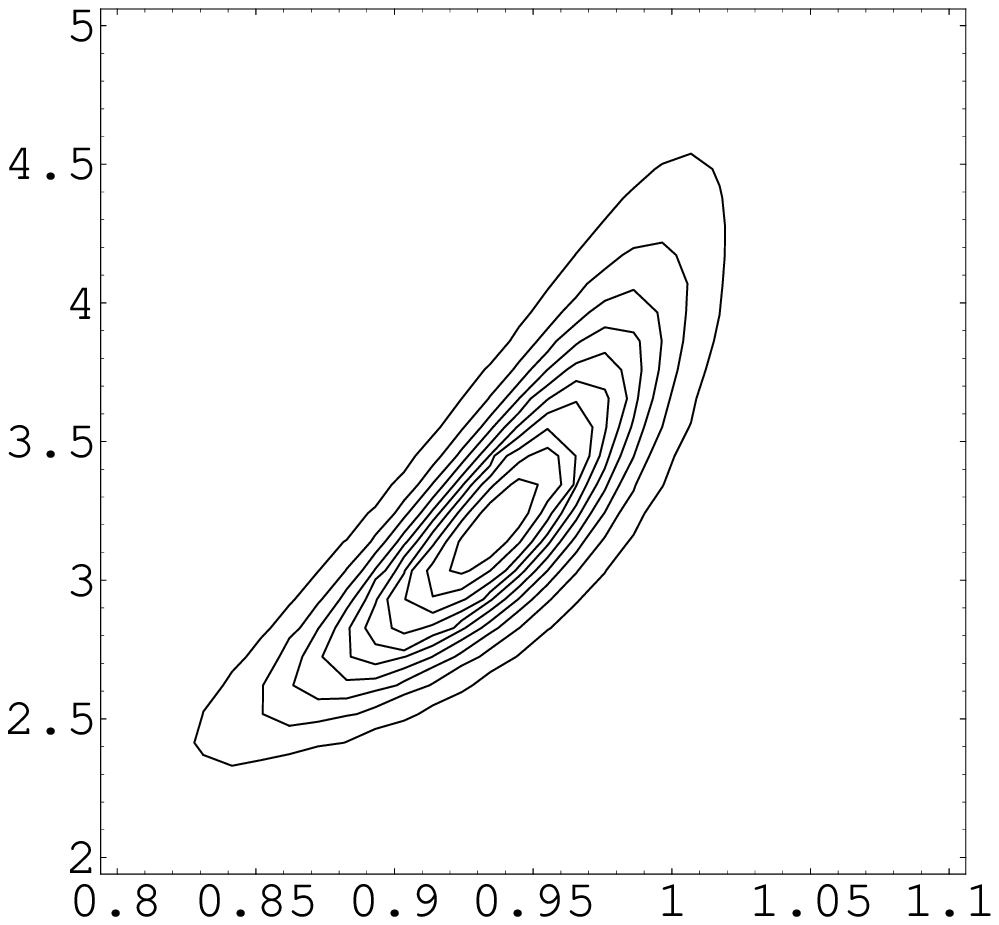}}}
 \put(0,0){\makebox(325,180)
          {\includegraphics{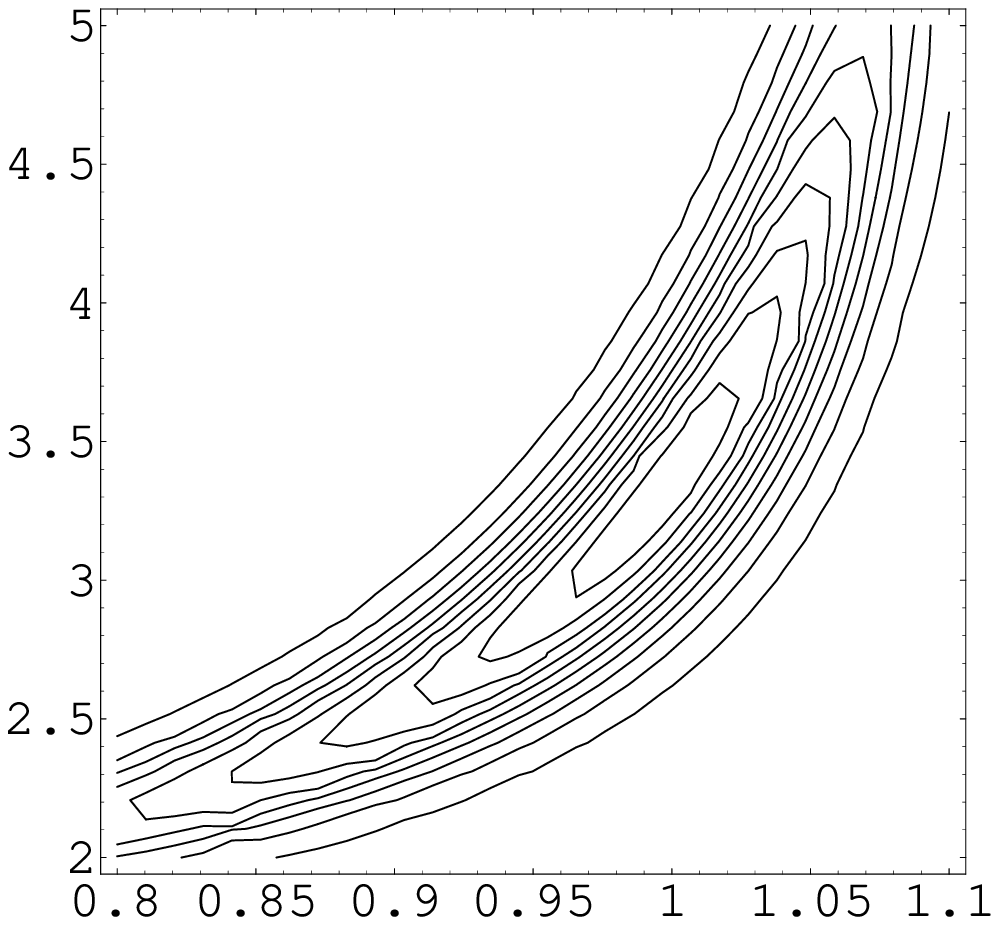}}}
\put(165.00,65.00){\makebox(0,0)[lc]{$\lambda_2$}}
\put(-20.00,65.00){\makebox(0,0)[lc]{$\lambda_2$}}
\put(70.00,-10.00){\makebox(0,0)[lc]{$\lambda_1$}}
\put(255.00,-10.00){\makebox(0,0)[lc]{$\lambda_1$}}
\end{picture}
\end{center}
\caption{
Contour plots of the posterior $P(\lambda_1,\lambda_2|D_{50})$.
Left: posterior corresponding to the MaxEnt sampling experiment
of Figure \ref{fig_multitq}. Right: posterior as obtained after equidistantly
sampling 50 data points between $t=0$ and $t_{max}=1.5$.
The chosen value of $t_{max}$ represents the optimal 
if one aims at highly accurate estimates of the parameters $\lambda$.
In contrast to monoexponential sampling,
however, the optimal choice of $t_{max}$ hardly
depends on the values of $\lambda$ for intermediate values
of $t_{max}$
}
\label{fig_multipost} 
\end{figure}

In Figure \ref{fig_multipost} we show the resulting posterior
on $\lambda$ (with $A_i$ and $\sigma$ treated as nuisance parameters)
in order to compare with equidistant sampling. 
As can be seen, sequential MaxEnt sampling leads to 
lower error bars in the estimates of $\lambda$. 
In particular the width of the posterior in the large decay constant
$\lambda_2$ is considerably larger for equidistant sampling.
For signals of the form (\ref{porous}) this is exactly what one
would expect since the term with the smaller decay constant $\lambda_1$
is sampled more frequently than the term decaying with $\lambda_2$.
For sequential MaxEnt sampling, however, 
the latter term is sampled more frequently,
cf.\ Figure \ref{fig_multitq}. 

Despite the good performance of sequential MaxEnt sampling it has to
be stressed that it becomes impracticable in its current form
for $n_{max} > 2$.
For a low number $N$ of experiments it will be difficult to find
evidence for the decay constant $\lambda_{n_{max}}$ in (\ref{porous}).
A low number of experiments $N$
will result in a degeneracy in the estimates of some of the
$\lambda_i$ leading to divergences in the posterior. The remedy here
would be to incorporate model selection as described in 
\cite{Bretthorst88} into the algorithm. This would probably result
in an algorithm which starts with the hypothesis $n_{max}=1$ for
low $N$ and increases $n_{max}$ as soon as the relative
probability for such a hypothesis is larger. 

\section{Summary\label{secSummary}}

Starting from the question at what value of the control parameter one should
perform the next experiment after having performed $N$ measurements, we have
analytically derived a sequential sampling procedure which maximizes  the
information on the parameters to be estimated.
Looking for the query $t_{N+1}$ that leads to a maximum information 
gain (minimization of the posterior
entropy) led us to the conclusion that one should find $t_q$ that {\em
maximizes} the entropy of the predictive distribution. The reason for this
is that $t_q$ is the value of the control
parameter for which the prediction is least secure.

We have applied the constructed MaxEnt sampling procedure  to the fitting
of exponential signals. We found that the resulting queries oscillate
between the time constants present in the signal.  The width of the
distribution of queries around the time constants 
depends on the variance of noise by which the
signal is corrupted.

For purely monoexponential signals we have compared sequential MaxEnt sampling
with equidistant sampling and found a considerably better accuracy
of the estimates of MaxEnt sampling. The application of sequential
MaxEnt sampling to 
multiexponential signals is of particular interest because
of their frequent appearance in science and the difficulties associated
with the inverse Laplace transform. We have motivated that  proper 
sampling helps to perform the inversion task. We have exemplified that
sequential MaxEnt sampling is capable of tracking the spectrum quickly. 

However, so far we have applied the algorithm only to an assumed
constant number of exponentially decaying terms.  This is a bit unrealistic,
since typically there would
be infinitely many such terms, like the NMR signals mentioned in Section
\ref{secMultiexp}. Moreover, the assumption of a large fixed number
of exponential terms in the signal leads to degeneracies and hence divergences
in the posterior. We hope that these divergences can be removed by
hypothesizing a small number $n_{max}$ of exponential terms at the onset of 
the experiment and increasing this number whenever the relative
probability favors the hypothesis of a larger value of $n_{max}$.

Finally we note that
due to the nonlinear dependence of the model functions on the decay constants
$\lambda$ results could only be obtained numerically. 
For models completely
linear in the model parameters it is possible to do the 
calculation of query times $t_q$ analytically \cite{unpublished}, however. 
For such a case, the result is rather trivial,  leading to $t_q
\to \infty$.

\medskip
The authors are grateful to S.\ Rabbani and C.\ Mendon\c{c}a
for stimulating discussions. 
P.R.\ acknowledges the hospitality extended to him at Universidade de S\~ao
Paulo where this work has been initiated.

\end{document}